\documentclass[11pt]{iopart}

\usepackage{newtxtext}

\makeatletter
\renewenvironment{abstract}
  {\par\noindent{\bfseries Abstract}\par
   \noindent}
  {\par}
\makeatother

\usepackage{iopams}

\usepackage{graphicx}
\usepackage{color}
\usepackage[normalem]{ulem}

\usepackage{graphicx}
\usepackage{cite}

\begin{document}

\setcounter{footnote}{0}
\renewcommand{\thefootnote}{\fnsymbol{footnote}}

\title{On Lorentz Variability of Magnetically Dominated Relativistic Outflows}

\author{V I Berezhiani$^{1,2}$, N L Shatashvili$^{2,3}$ and 
A G Tevzadze$^{3,4,*}$\footnotemark[1]}

\address{$^1$ School of Physics, Free University of Tbilisi, Tbilisi 0159, Georgia}
\address{$^2$ Andronikashvili Institute of Physics, TSU, Tbilisi 0177, Georgia}
\address{$^3$ Department of Physics, Faculty of Exact and Natural Sciences,
Ivane Javakhishvili Tbilisi State University, Tbilisi 0179, Georgia}
\address{$^4$ Evgeni Kharadze Georgian National Astrophysical Observatory, Abastumani 0301, Georgia}

\renewcommand{\thefootnote}{\fnsymbol{footnote}}

\newcommand{\keywords}[1]{%
  \par\noindent\textbf{Keywords:} #1\par
}

\begin{abstract}
We show that magnetized relativistic outflows can exhibit a
relativistic effect in which Lorentz transformation maps spatially extended magnetic field structure into apparent temporal variability in the observer’s frame.
Using a force-free Beltrami configurations as representative equilibria of magnetically
dominated outflows, we demonstrate that Lorentz mapping of stationary helical magnetic field
produces quasi-periodic modulation of observable electromagnetic signatures, without invoking
intrinsic plasma variability. 
The same mechanism naturally generates temporal evolution of the observed polarization properties, producing smooth, generally aperiodic polarization-angle swings in the observer's frame.
This effect may be described as an aberration of force-free magnetic fields under Lorentz transformation.  The characteristic frequency of the time variability is determined by the helical wave-number of the magnetic field, 
the viewing angle, and the bulk Lorentz factor of the jet outflow, and scales linearly 
with $\Gamma$. 
This establishes a purely kinematic relativistic origin of variability and introduces
the concept of magnetic Lorentz seismology:
the inference of magnetic field structure in relativistic outflows directly
from observed temporal variability.

\end{abstract}
~\\

\section{Introduction}

Magnetic fields in relativistic plasma flows often exhibit behavior far outside
the regimes accessible in laboratory experiments. Such conditions naturally occur in
relativistic jets, which provide the most prominent natural laboratories for this physics:
they host magnetically dominated regions with extremely high bulk Lorentz factors,
in which the field controls both the flow structure and its variability. Such jets
are observed in blazars, quasars, gamma-ray bursts, active galaxies, and X-ray binaries,
offering multiple contexts in which these extreme magnetic field regimes occur.
{\color{black} Furthermore, recent theoretical and observational developments have reinforced the importance of magnetic fields in shaping the dynamics, emission, and variability of relativistic jets \cite{Bromberg2024}. }
In these environments, the field may dominate the plasma energy density and thereby
enforce a force-free configuration (see, e.g., \cite{BZ1977,K2007}).

Relativistic jets are commonly associated with supermassive black holes and other compact objects, where highly energetic plasma outflows originate. Understanding the physical conditions in the emitting regions is therefore essential for the reliable interpretation of observations. These jets extend to distances far from the central compact object, typically reaching parsec scales, which correspond to $\sim 10^{4}-10^{5} r_g$, where $r_g$ is the gravitational radius of the central object \cite{Blandford2019}. At such distances, spacetime curvature is negligible and the dynamics is well described by Minkowski spacetime. General relativistic effects, such as gravitational lensing or frame dragging, are confined to the immediate vicinity of the central object (within a few gravitational radii) and do not influence the kinematic mapping. Emission produced in these extended regions is therefore insensitive to general relativistic effects and can be consistently analyzed within the framework of special relativity.

Interpreting the observational features of relativistic plasma flows is highly nontrivial 
due to the fact that
mapping from rest-frame physics to the observer frame is profoundly altered
by Lorentz transformations. These distortions grow rapidly with bulk Lorentz factor,
so that in super-relativistic flows a number of profound effects can be identified.
These include patterns moving with superluminal speeds \cite{R1966,C1977},
relativistic beaming \cite{BK1979},
Doppler and $\Gamma$-boosting effects \cite{BBR1984},
and relativistic polarization-angle variations \cite{LPG2005,L2022}.
A theory of magnetic fields frozen in such flows must therefore incorporate both the
intrinsic structure and variations of the field, as well as their observational
signatures in super-relativistic regimes. 
Observed variability in relativistic jets is commonly attributed to shocks, turbulence, magnetic reconnection, Doppler boosting, lighthouse effects, or geometric jet precession. These mechanisms invoke either intrinsic plasma dynamics or temporal changes in the orientation of the emitting region relative to the observer. Here, by contrast, we intentionally consider a stationary magnetic configuration in order to isolate the consequences of relativistic Lorentz mapping and quantify the observational bias it may introduce when inferring the intrinsic properties of relativistic outflows. The resulting variability emerges solely from the Lorentz transformation of a spatially structured magnetic field, which maps stationary magnetic structure in the co-moving frame into apparent temporal variability in the observer's frame.

In relativistic plasma flows, a consistent description of magnetic fields must be formulated in terms of covariant electromagnetic quantities. While the concept of magnetic field lines is often employed as a visual aid in astrophysical and plasma applications, such representations do not constitute a fundamental description in relativistic regimes, particularly when the field structure is highly inhomogeneous and intrinsically relativistic, as in the case of relativistic jets. In such systems, the physically meaningful quantities are the components of the electromagnetic field tensor and their transformation between reference frames. 
Accordingly, we adopt a fully covariant formulation based on the electromagnetic field tensor $F^{\mu\nu}$. The analysis does not rely on any notion of field-line motion: all results follow directly from Lorentz transformations of the electromagnetic field tensor and the corresponding mapping of spacetime coordinates.

While the Lorentz transformation of electromagnetic fields is formally straightforward, its implications for spatially extended and structured field configurations are nontrivial. In such systems, the mapping between spatial structure in the co-moving frame and spacetime coordinates in the observer frame can give rise to observable effects that have no direct analogue in point-like systems. This is illustrated by well-known relativistic effects: in addition to apparent superluminal motion, a related example is the Penrose-Terrell rotation \cite{Penrose1959,Terrell1959}, recently demonstrated in laboratory experiments using ultrafast optical techniques \cite{Hornof2025}.

In the present work we show that a magnetic field that is strictly stationary
in the co-moving frame of a relativistic jet can nonetheless produce apparent
periodic variations in the electromagnetic emission observed by an external
observer. This effect arises solely from Lorentz mapping of spatially extended
magnetic structure, linking it directly to observable temporal variability.
To model the underlying field structure, we adopt a force-free Beltrami
configuration as a representative equilibrium of highly magnetized relativistic
outflows. Within this framework, we derive the characteristic frequency of the
variability and demonstrate that it follows purely from relativistic kinematics
and projection effects. Consequently, the observed time variability provides
a direct probe of the underlying magnetic field structure near the jet axis
in super-relativistic flows.

\section{Model and Analysis} 

To explore how spatially extended stationary magnetic structures 
behave under relativistic transformations, we develop a covariant formalism that 
relates their representation in the co-moving and observer frames. For this purpose
we adopt relativistic
jet geometry shown in Fig. 1.
The coordinate system $(X,Y,Z)$ defines the observer's lab frame,
with distant observer located along the $Z$-axis.
The jet lies in the plane normal to the $X$-axis and is inclined by an angle
$\theta$ to the line of sight.
The primed spatial coordinates $(X^\prime,Y^\prime,Z^\prime)$
correspond to the local co-moving frame of the jet outflow,
which moves at relativistic velocity relative to the observer.
The coordinate system $X^{\prime \prime},Y^{\prime \prime},Z^{\prime \prime}$
specifies yet another co-moving frame,
with its $Z^{\prime \prime}$ axis aligned with the direction of the jet flow.
Lorentz transformations are then applied to relate quantities between the two frames,
enabling us to compute the corresponding observable features in the lab frame.
Note, that no assumptions are made about geometric alignment or the physical interpretation of magnetic field lines, which are frame-dependent under Lorentz transformations. In this setup, the only physically meaningful geometrical parameter is the angle $\theta$ between the observer’s line of sight and the jet bulk velocity, as it directly enters the Lorentz transformation and determines the relative orientation of the reference frames.

\begin{figure}[t]
\centering
\begin{minipage}{0.4\textwidth}
    \includegraphics[width=\linewidth]{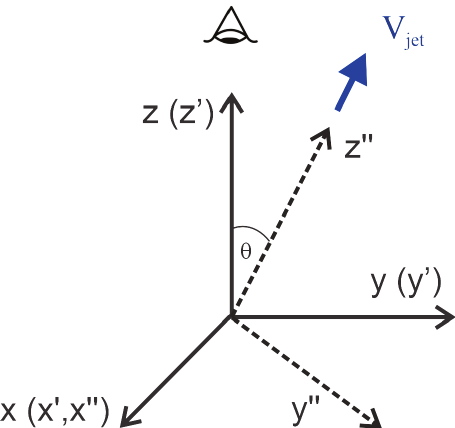}
\end{minipage}
\hspace{-0.05 \textwidth}
\begin{minipage}{0.60\textwidth}
\caption{\label{fig1}
Schematic illustration of the coordinate systems used in the analysis.
The system $(X,Y,Z)$ defines the observer’s lab frame, with the observer
(line of sight) along the $Z$-axis. The jet lies in the plane perpendicular
to the $X$-axis and is inclined by an angle $\theta$ to the line of sight.
The primed frames $(X^\prime,Y^\prime,Z^\prime)$ and
$(X^{\prime\prime},Y^{\prime\prime},Z^{\prime\prime})$ are co-moving
coordinate systems, with the $Z$ and $Z^\prime$ axes aligned with the line
of sight and the $Z^{\prime\prime}$ axis aligned with the jet plasma flow
direction, respectively.}
\end{minipage}
\end{figure}

The electromagnetic field in the co-moving frame is described by the
field tensor $F^{\prime\mu\nu}$, defined at the co-moving spacetime
coordinates $x^{\prime\delta}$. The corresponding field in the observer
frame is obtained through the Lorentz transformation of the tensor:
\begin{equation}
F^{\mu\nu}(x^\delta) = \Lambda^{\mu}{}_{\alpha}(-\boldsymbol{\beta}) 
\Lambda^{\nu}{}_{\beta}(-\boldsymbol{\beta}) F^{\prime\alpha\beta}(x^{\prime\delta}) ~.
\end{equation}
Here $\Lambda^{\mu}{}_{\alpha}(\boldsymbol{\beta})$ denotes the
Lorentz transformation matrix from the observer's into the co-moving frame,
which moves with the outflow velocity $\boldsymbol{\beta} = \boldsymbol{v}/c$
and the inverse Lorentz transformation is implemented by reversing the
velocity sign ($-\boldsymbol{\beta}$).
Since the co-moving tensor is defined in co-moving coordinates, the
transformation must be also applied consistently to the spacetime arguments as well:
\begin{equation}
x^{\prime\delta}
= {\Lambda^{\delta}}{}_{\eta}(\boldsymbol{\beta})\,x^{\eta}.
\end{equation}
Assuming that transverse motions perpendicular to the jet axis are non-relativistic
and negligible compared to the jet outflow velocity $|\boldsymbol{v} - \boldsymbol{v}_{\rm jet}|\ll |\boldsymbol{v}_{\rm jet}|$, we ignore their contribution to the total bulk Lorentz factor and parameterize the jet velocity as:
\begin{eqnarray}
\beta_x &=& 0 ~,~ \beta_y = \beta \sin(\theta) ~,~ \beta_z = \beta \cos(\theta) ~, \\
\Gamma &=& 1/\sqrt{1-v_{\rm jet}^2/c^2} = 1/\sqrt{1- \beta^2} ~.
\end{eqnarray}
Adopting the Minkowski metric tensor with signature ($+$,$-$,$-$,$-$),
the Lorentz transformation matrix for a boost with
$\boldsymbol{\beta}=(0,\beta_y,\beta_z)$ takes the form:
{\color{black}
\begin{equation}
\Lambda^{\mu}{}_{\nu}(\boldsymbol{\beta}) =
\left[
\begin{array}{cccc}
\Gamma & -\Gamma \boldsymbol{\beta} \\[6pt]
-\Gamma \boldsymbol{\beta} & \delta_{ij} + (\Gamma-1) {\beta_i \beta_j / \beta^2} \\[6pt]
\end{array}
\right].
\end{equation}
where $\delta_{ij}$ is the Kronecker delta and $i,j$ are spatial indices.}

For simplicity, we assume that the large scale electric field in the co-moving
frame is absent (${\bf E^\prime} = 0$), and the equilibrium
magnetic field obeys the Beltrami condition:
\begin{equation}
\nabla \times {\bf B^\prime} = \lambda {\bf B^\prime} ~,
\end{equation}
where $\lambda$ is constant Beltrami parameter.
A Beltrami vector field -- a particular type of equilibrium state in fluids and plasmas 
\cite{CK1957,low} -- has wide applicability as one of the simplest models of vortex structures 
observed in nature. The Beltrami condition, requiring alignment of the flow with its vorticity, 
enforces a homogeneous distribution of the total energy density (a Bernoulli condition), 
a nontrivial property given the presence of finite helicity. In the study of force-free magnetic fields,
\cite{Woltier} emphasized the significance of the Beltrami field in relation to helicity and 
relaxation phenomena: helicity was invoked as a constraint in minimizing the magnetic energy, 
yielding the Beltrami condition as the Euler–Lagrange equation, which takes the form of 
an eigenvalue problem for the {\it curl} operator. Subsequently, \cite{Taylor} proposed the 
concept of a ``relaxed state'' as the energy minimum under a rugged invariant; 
he postulated that helicity is approximately conserved while plasma turbulence 
dissipates magnetic energy. The resulting Taylor-relaxed states (i.e., Beltrami magnetic fields)
have been observed in various laboratory experiments \cite{Taylor} as well as in some 
astrophysical systems \cite{kusano}. 
The helicity constraint implies finite vorticity in the relaxed state, leading to nontrivial topological
properties of magnetic field lines \cite{Moffatt}. Several generalizations have been proposed, including 
the inclusion of cross helicity as an additional constraint when the flow is parallel to the magnetic field;
two-fluid (Hall MHD) formulations that elucidate the coupling between flow and magnetic field in terms 
of canonical vorticities \cite{MY1998}; and simultaneous ion (or multi-ion) and electron Beltrami conditions,
yielding double- or multi-Beltrami fields with applications in both laboratory and astrophysical plasmas
\cite{mnsy,osym,RT1,SMB2T,SSMD}. 
Indeed, it is known that Beltrami magnetic fields, as an exact force-free solutions,
provide a mathematically transparent and physically consistent model for the
large-scale ordered magnetic fields typical of relativistic jets \cite{Katsadze} 
and references there in).

In magnetically dominated regions of relativistic jets, where the plasma beta is extremely low, magnetic relaxation processes (Taylor relaxation) are expected to drive the field toward minimum energy states constrained by magnetic helicity, naturally leading to force free Beltrami configurations. Furthermore, provided that the global jet properties evolve on time scales much longer than the characteristic internal dynamical time scales, the resulting magnetic structure may be regarded as quasi-stationary in the co-moving frame. The present work is restricted to this regime in order to isolate the effects of relativistic Lorentz mapping. Such quasi-stationary configurations can persist as long as the global jet boundary conditions evolve on timescales longer than the internal Alfvén crossing time.
{\color{black}
Moreover, numerical simulations of relativistic magnetized jets indicate that large scale ordered helical magnetic fields can survive over many dynamical times even in the presence of turbulence and magnetic dissipation. While local instabilities and magnetic reconnection may distort the field structure, the global magnetic topology is often preserved, supporting the use of force-free Beltrami configurations as idealized models of the large scale magnetic structure \cite{Mizuno2012,Mizuno2009}. The present work is intended as a proof-of-principle study that isolates the consequences of relativistic Lorentz mapping in such ordered configurations.}

To model a force-free magnetic field in the jet, we employ a local approximation to 
the cylindrical Chandrasekhar-Kendall (CK) modes \cite{CK1957,Y1991}, when jet direction 
($Z^{\prime \prime}$-axis) coincides with the the cylindrical symmetry axis of these modes:
\begin{eqnarray}
B_r^{\prime \prime} &=& -{k \over \alpha} B_0 J_1(\alpha r^{\prime \prime}) \cos(k z^{\prime \prime}) \approx 0 ~, \\
B_\phi^{\prime \prime} &=& {\lambda \over \alpha} B_0 J_1(\alpha r^{\prime \prime}) \sin(k z^{\prime \prime}) \approx B_0 J_1(\alpha r^{\prime \prime}) \sin(k z^{\prime \prime})  ~, \\
B_z^{\prime \prime} &=& B_0 J_0(\alpha r^{\prime \prime}) \sin(k z^{\prime \prime}) ~,
\end{eqnarray}
where ${r^{\prime \prime}} = \sqrt{{x^{\prime \prime}}^2 + {y^{\prime \prime}}^2}$ is the radial coordinate normal to the outflow and $\lambda^2 = \alpha^2 + k^2$ is the Beltrami helical parameter that describes pitch angle of the force-free configuration. In the local co-moving Cartesian frame:
\begin{eqnarray}
B_x^{\prime \prime} &=& \frac{x^{\prime \prime}}{r^{\prime \prime}} B_r^{\prime \prime} - \frac{y^{\prime \prime}}{r^{\prime \prime}} B_\phi^{\prime \prime} ~, \\
B_y^{\prime \prime} &=& {y^{\prime \prime} \over r^{\prime \prime}}  B_r^{\prime \prime} + {x^{\prime \prime} \over r^{\prime \prime}}  B_\phi^{\prime \prime}  ~,
\end{eqnarray}
and near the jet axis, where $\alpha r^{\prime \prime} \ll 1$, the Bessel functions may be well approximated by:
$$
J_0(\alpha r^{\prime \prime}) \approx 1  ~,~~ J_1(\alpha r^{\prime \prime}) \approx {\alpha r^{\prime \prime} / 2} ~.
$$
To account for the misalignment of the jet axis and observer's line of sight, we transform CK modes into the primed co-moving frame by using the counterclockwise spatial rotation around the $X$-axis:
\begin{equation}
{\boldsymbol B}^\prime = {\cal R}_x(-\theta) {\boldsymbol B}^{\prime \prime} ~,
\end{equation}
with the rotation matrix:
\begin{equation}
{\cal R}_x(\theta) =
\left(\begin{array}{ccc}
1 & 0 & 0 \\ 0 & \cos(\theta) & -\sin(\theta) \\ 0 & \sin(\theta) & \cos(\theta)
\end{array}\right) ~.
\end{equation}
Supplementing this with corresponding space coordinate rotation:
\begin{equation}
{\boldsymbol r}^{\prime \prime} = {\cal R}_x(\theta) {\boldsymbol r}^{\prime}
\end{equation}
we derive stationary magnetic field configuration depending on local co-moving coordinates as follows:
\begin{eqnarray}
B_x^\prime &=& - B_0 {\alpha \over 2} l_T^\prime \sin(k l_H^\prime) ~, \\
B_y^\prime &=& B_0 \left[ \sin(\theta) + {\alpha \over 2} \cos(\theta) x^\prime \right] \sin(k l_H^\prime) ~, \\
B_z^\prime &=& B_0 \left[ \cos(\theta) - {\alpha \over 2} \sin(\theta) x^\prime \right] \sin(k l_H^\prime) ~,
\end{eqnarray}
where we have introduced effective helical coordinate $l_H^\prime$ and orthogonal transverse coordinate $l_T^\prime$ as follows:
\begin{eqnarray}
l_H^\prime &=& \cos(\theta) z^\prime  - \sin(\theta) y^\prime ~, \\
l_T^\prime &=&  \cos(\theta) y^\prime + \sin(\theta) z^\prime ~.
\end{eqnarray}

To recover properties of such stationary helical magnetic fields in the observer's frame, we must apply Lorentz transformation of the electromagnetic field tensor, followed by transformation of 
spacetime arguments.
{\color{black} A detailed derivation of the CK field in the observer’s frame, resulting in explicit expressions for the electromagnetic field components as functions of the observer frame spacetime coordinates, is presented in the Appendix A. It seems that the magnetic field in the observer’s frame takes the following form:
\begin{eqnarray}
B_x({\bf r},t) &=& - B_0 {\alpha \Gamma \over 2} L_T({\bf r},t) \sin(\Phi_H({\bf r},t)) ~, \\
B_y({\bf r},t) &=& B_0 \left( \sin(\theta) + {\alpha \Gamma  \over 2} \cos(\theta) x \right)
\sin(\Phi_H({\bf r},t)) ~, \\
B_z({\bf r},t) &=& B_0 \left( \cos(\theta) - {\alpha \Gamma \over 2} \sin(\theta) x \right)
\sin(\Phi_H({\bf r},t)) ~.
\end{eqnarray}
where $\Phi_H({\bf r},t)$ is the helical phase and $L_T({\bf r},t)$ is the transverse coordinate, whose explicit forms are to be found in the Appendix, Eqs. (A.18) and (A.19).
}

Although the field is assumed to be stationary and purely magnetic in the comoving frame,
Lorentz transformation necessarily give rise to electric field components in the
observer’s frame through relativistic $E$–$B$ mixing. We do not explicitly display these
electric field components, as they play no dynamical role in the variability discussed here,
which arises entirely from Lorentz mapping of magnetic field structure.
{\color{black}
The electric field components nevertheless remain an essential part of the covariant electromagnetic structure and are presented explicitly in Appendix A for completeness. Since the synchrotron emission and polarization signatures considered in this work are determined primarily by the projected magnetic field geometry, the transformed electric field does not introduce an independent variability timescale or modify the characteristic frequency associated with the Lorentz mapping of the magnetic structure. Consequently, the principal observational signatures discussed below are governed by the magnetic field components.
}

Interestingly, helical and transverse coordinates become time-dependent in the observer's frame
due to the misalignment of the observer and jet axis and relativistic $\Gamma$ boost:
{\color{black}
\begin{eqnarray}
\frac{\partial}{\partial t} \Phi_H({\bf r},t)  &=& -\beta c \Gamma \cos(2\theta)\,k ~, \\
\frac{\partial}{\partial t} L_T({\bf r},t) &=& -\beta c \Gamma \sin(2\theta) ~. 
\end{eqnarray}
}
The explicit time dependence appearing in Eqs. (23-24) implies corresponding temporal variability
of the magnetic field components in the observer’s frame (see Eqs. 20-22).
It seems, therefore, that a magnetic field which is stationary in the co-moving frame
of a relativistic jet can nonetheless exhibit \emph{apparent time variability in the observer’s frame}
as a direct consequence of Lorentz transformations. In the present Letter we demonstrate
this effect explicitly, showing that force-free helical magnetic fields, although intrinsically
steady in the co-moving frame, can produce apparent time variability solely due to their
spatial structure, without invoking any intrinsic temporal evolution of the field.\\

\section{Observational Signatures}

To analyze the observational properties of such magnetic fields, we emphasize that
emission from relativistic jets is dominated by synchrotron radiation,
which is produced by the magnetic field component perpendicular to the line of sight.
The synchrotron emissivity therefore depends only on the projection of the magnetic field
onto the sky plane.
For unresolved or marginally resolved sources, the internal magnetic structure of the jet
cannot be accessed directly and must be inferred solely from observable quantities.
As a result, all measurable characteristics, including the total flux, polarization, and variability,
depend entirely on the perpendicular magnetic field component, whose geometry and
apparent evolution are therefore the primary physical properties probed by observations
(see, e.g., \cite{GS1965,RL1979}).
For simple qualitative estimates, we therefore evaluate the observed flux using
the apparent magnetic field component perpendicular to the line of sight as:
\begin{equation}
F_\nu \propto B_\perp^2 = B_x^2 + B_y^2 ~.
\end{equation}
{\color{black}

We emphasize that Eq. (25) provides only a simplified estimate of the variability induced by Lorentz mapping of the magnetic structure. In a realistic synchrotron source, the observed emissivity depends not only on the perpendicular magnetic field component, but also on the electron energy distribution, radiative cooling, pitch angle distribution, and relativistic Doppler boosting (see. e.g., \cite{Blandford2019,L2022}). In optically thin synchrotron emission with a power-law electron distribution, $N(E)\propto E^{-p}$, the emissivity generally scales as $F_\nu \propto B_\perp^{(p+1)/2}$. Typical relativistic jet spectra correspond to $p \sim 2 - 3$, yielding a dependence close to $B_\perp^{1.5 - 2}$. The present treatment therefore captures the dominant magnetic contribution to the variability while intentionally isolating the purely kinematic effect associated with Lorentz transformation of a stationary spatially structured magnetic field. Although this simplified approximation may somewhat overestimate the modulation amplitude, it preserves the essential relativistic mechanism by which stationary spatial magnetic structure is mapped into apparent temporal variability in the observer frame.
}

Hence, we may estimate the time variability of the radiation flux using our Beltrami CK model
as follows:
\begin{equation}
F_\nu(\theta,t) = F_0(\theta,t) \sin^2(\Phi_H(\mathbf{r},t) ) ~,
\end{equation}
where aperiodic part can be described by:
\begin{equation}
F_0(\theta,t) = B_0^2 {\alpha^2 \Gamma^2 \over 4}
\left[ L_T^2(t) + \left( {2 \over \alpha \Gamma} \sin(\theta) + \cos(\theta) x \right)^2 \right] ~.
\end{equation}
We may now estimate the characteristic frequency temporal variability of the observed flux for small viewing angles ($\theta \ll 1$) and relativistic outflows ($\beta \simeq 1$) as follows:
\begin{equation}
\omega_{\rm obs} = 2 k c \Gamma ~.
\end{equation}

The expression for $\omega_{\rm obs}$ highlights a key conceptual result of this work.
Although the magnetic field configuration is strictly stationary in the co-moving frame,
its intrinsic spatial helicity give rise to a well-defined temporal frequency in the observer’s frame.
In particular, a perfectly stationary Beltrami magnetic field in the co-moving frame produces an intrinsically periodic light curve in the observer frame, with the period set entirely by the bulk Lorentz factor $\Gamma$,the viewing angle $\theta$ and the parameter $k$ that describes the vertical variability of the helical magnetic field structure. We may identify this effect as \emph{relativistic aberration} of helical magnetic fields produced by Lorentz transformation.

Importantly, $\omega_{\rm obs}$ is not associated with any propagating wave or
plasma instability. Instead, it reflects a purely kinematic mapping of spatial structure
into apparent temporal variability through relativistic transformations.
This establishes a previously unexplored channel for variability in relativistic jets,
linking observed timescales directly to the helical structure of force-free magnetic fields.

The derived result provides an explicit and testable scaling, implying that measurements
of variability timescales can directly constrain the magnetic pitch helical structure of relativistic magnetized jets, even when the source is unresolved.

{\color{black}

We may now estimate periodic variations of the polarization properties of synchrotron radiation produced by Lorentz variability of the magnetic structure. Since synchrotron polarization is determined by the projected magnetic-field geometry, the Lorentz-induced time dependence of the observer-frame magnetic field components naturally leads to temporal modulation of the Stokes parameters. For a qualitative estimate, the corresponding Stokes parameters may be written as
\begin{equation}
I \propto B_x^2+B_y^2 ~,~~~ Q \propto B_y^2-B_x^2 ~,~~~ U \propto -2B_xB_y ~.
\end{equation}
Here we neglect the circular polarization ($V\simeq0$), since optically thin synchrotron emission from ultrarelativistic electrons is expected to be predominantly linearly polarized, while circular polarization may additionally require propagation or plasma conversion effects not included in the present model.
{\color{black} It should be noted that propagation effects such as Faraday rotation, depolarization, and line-of-sight averaging may modify the detailed observational signatures of the predicted polarization variations, but are not expected to eliminate the underlying kinematic mechanism associated with Lorentz mapping of a stationary spatially structured magnetic field.}

The linear polarization angle in the observer's frame may then be estimated as:
\begin{equation}
\chi(\theta,t) = \frac{1}{2}\arctan\left(\frac{U}{Q}\right) =
\arctan\left( \frac{\alpha \Gamma L_T(t)} {2\sin\theta + \alpha \Gamma \cos\theta\,x} \right) ~.
\end{equation}
Unlike the flux variability, which is intrinsically periodic, the polarization angle evolution is generally not strictly periodic. Instead, the Lorentz-transformed transverse coordinate enters the polarization angle through an arctangent dependence, leading to continuous polarization-angle swings or rotations in the observer’s frame. Thus, the present mechanism naturally predicts periodic flux modulation accompanied by secular or quasi-monotonic evolution of the linear polarization angle.

For small viewing angles ($\theta \ll 1$), the polarization-angle evolution may become significantly enhanced, yielding the characteristic polarization angle swing rate:
\begin{equation}
\frac{d\chi}{dt} \propto \alpha c \Gamma^2 ~.
\end{equation}

Thus, the longitudinal (helical) wave number $k$ of the Beltrami magnetic structure determines the characteristic frequency of the intensity variability, whereas the transverse (radial) wave number $\alpha$ controls the rate of the aperiodic polarization angle swings.
Thus, the CK relativistic aberration model naturally predicts rapid flux variability accompanied by relatively smooth polarization-angle swings.

{\color{black}
We do not analyze the detailed angular dependence of the variability and polarization signatures, since synchrotron and curvature radiation from ultrarelativistic outflows are strongly beamed within a narrow cone of opening angle $\theta \sim 1/\Gamma$, while the emitted intensity decreases rapidly for larger viewing angles. Consequently, a broad parameter study over $\theta$ would extend far beyond the narrow angular range that contributes significantly to the observed emission and may therefore obscure the physically relevant regime considered in the present work.
Extending this framework to more general magnetic configurations may provide a systematic way to infer magnetic field structure from variability data.

More generally, the principal objective of this work is to establish the existence of a previously unexplored relativistic effect arising from Lorentz mapping of stationary spatially structured magnetic fields. Extending this framework to more general magnetic configurations may provide a systematic way to infer magnetic field structure from variability data.
}

\begin{figure}
\centering
\begin{minipage}{0.55\textwidth}
\centering
\includegraphics[width=\linewidth]{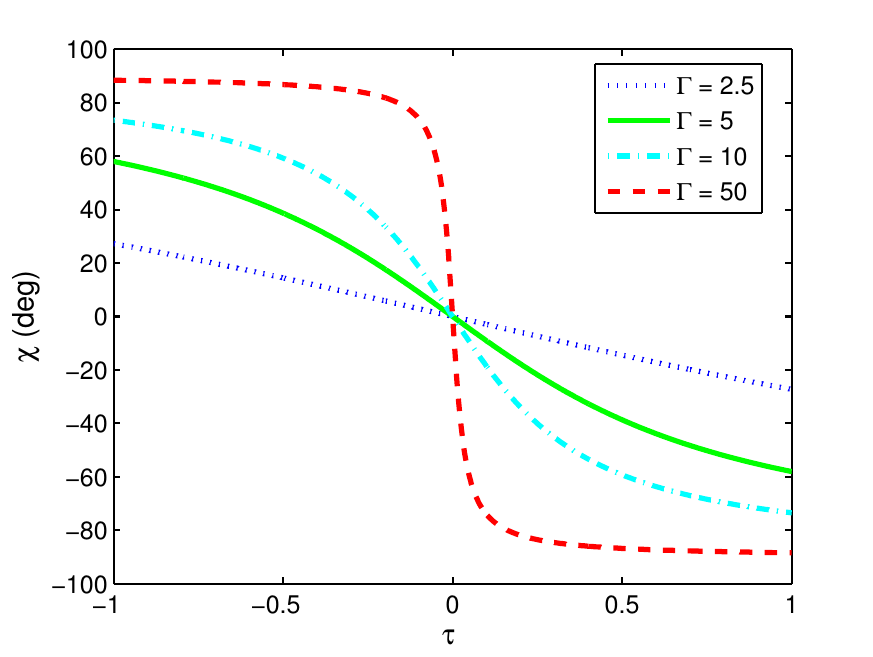}
\end{minipage}
\hspace{-0.18\textwidth}
\begin{minipage}{0.60\textwidth}
\caption{\label{fig2}
Evolution of the linear polarization angle $\chi$ as a function of the dimensionless observer time $\tau = k c t$ for different jet bulk Lorentz factors, $\Gamma=2$, $5$, $10$, and $50$. The parameters are $\alpha=1$, $\theta=1^\circ$, $y=z=0$, and a small transverse offset $x=0.1$. The characteristic timescale of the polarization swing decreases rapidly with increasing $\Gamma$. 
{\color{black}
The chosen parameters are representative of the small-angle regime relevant for ultrarelativistic outflows, where the proposed Lorentz-mapping effect is most pronounced. Increasing $\Gamma$ leads to progressively faster polarization swings, illustrating the relativistic bias toward rapid EVPA evolution in highly relativistic jets.
}}
\end{minipage}
\end{figure}

The present analysis is based on an idealized helical (Beltrami) configuration; more complex or turbulent magnetic structures may lead to broader or non-periodic variability patterns. Extending this framework to more general magnetic configurations may provide a systematic way to infer magnetic field structure from variability data.
Since the effect originates from Lorentz transformation of a spatially structured magnetic field rather than from any specific property of the CK solution, we expect the qualitative mechanism to persist for more general magnetic configurations, although the resulting variability may become broadband, irregular, or non-periodic. Extending this framework to more general magnetic configurations may provide a systematic way to infer magnetic field structure from variability data.

To illustrate the astrophysical implications of Eq.~(29), it is useful to express the variability timescale in terms of the characteristic helical scale of the magnetic field, $\lambda_H = 2\pi/k$. For numerical estimates, we restore the speed of light $c$, which was set to unity throughout the derivation. Equation (32) then takes the form:
\begin{equation}
T_{\rm obs} \simeq \frac{2 \pi}{\omega_{\rm obs}} = \frac{\lambda_H}{ 2 c \Gamma} ~.
\end{equation}
This relation shows that the observed timescale is determined entirely by the intrinsic magnetic-field scale and the bulk Lorentz factor of the outflow. 

Gamma-ray bursts (GRBs) provide the most extreme known relativistic outflows, with bulk Lorentz factors typically in the range $\Gamma\sim10^2-10^3$ (see, e.g., \cite{Zhang}). For compact magnetic structures with characteristic scales $\lambda_H \sim (10^{15}-10^{16})\,{\rm cm}$, Eq.~(32) predicts variability timescales of order $(10-10^3)\,{\rm s}$, corresponding to tens of seconds to tens of minutes. These values are comparable to the rapid variability commonly observed during the prompt-emission phase of GRBs.

Relativistic jets associated with blazars and other active galactic nuclei (AGN) typically exhibit Lorentz factors in the range $\Gamma\sim10-50$ (see, e.g., \cite{Blandford2019,Kim2024}). For the same magnetic scales, the proposed mechanism naturally produces variability on timescales ranging from several minutes to a few hours. Such timescales are comparable to the fastest variability observed in blazars \cite{Albert2007,MAGIC,Liodakis2025}.

Thus, measurements of variability timescales may provide direct constraints on the characteristic magnetic scale $\lambda_H$ and the bulk Lorentz factor $\Gamma$, opening a new avenue for inferring the underlying magnetic-field structure of unresolved relativistic jets through {magnetic Lorentz seismology}.
{\color{black}
Since the observed variability emerges from the interplay between the intrinsic magnetic structure and relativistic kinematics, distinct physical configurations may in some cases produce similar observational signatures. Future source-specific modeling, together with complementary observations, may help disentangle these degeneracies and establish magnetic Lorentz seismology as a practical tool for probing relativistic outflows.
Quantitative comparison with individual sources, however, requires source-specific modeling of the magnetic-field geometry, particle distribution, radiative transfer, and propagation effects. Such studies may further establish magnetic Lorentz seismology as a practical tool for probing the magnetic structure of relativistic outflows.
}

\section{Discussion and Conclusions}

In this work, we show that a stationary, spatially extended magnetic field structure in a relativistic plasma flow can be revealed as an apparent temporal variability in the observer frame as a direct consequence of Lorentz transformations. In this context derived results establish a previously unexplored mechanism in which Lorentz mapping of a Beltrami  fields into the observer’s frame gives rise to a quasi-periodic oscillation (QPO)–like variability and associated changes in observable emission properties,
including flux, polarization, and their temporal evolution. This effect arises entirely from the spatial
structure of the magnetic field and relativistic frame transformations, without any intrinsic dynamical
evolution of the magnetic field or plasma, and therefore represents a fundamentally kinematic origin
of variability in relativistic jets.

The linear scaling of the observed frequency with the bulk Lorentz factor implies
that this variability mechanism is strongly enhanced in super-relativistic jets. As a
result, observed quasi-periodic variability and polarization modulation may provide a
direct diagnostic of both the bulk Lorentz factor and the underlying magnetic field
structure, even in unresolved sources.
 
{\color{black} 
Observational tests of the proposed mechanism may be possible through
high-cadence multiwavelength monitoring of relativistic jets. In particular,
coordinated flux and polarization measurements, including radio and optical
polarimetry, VLBI observations of jet structure, and recent X-ray polarimetry
missions such as IXPE, may provide direct probes of the predicted connection
between variability timescales, polarization evolution, and the underlying
magnetic topology. Detection of periodic flux modulation accompanied by
systematic polarization-angle swings would provide a particularly distinctive
signature of the relativistic magnetic-structure mapping proposed here.

}

While Lorentz transformations of electromagnetic fields are well known, the nontrivial result demonstrated here is that a purely spatially structured, stationary magnetic field associated with a spatially extended source can give rise to a well-defined temporal frequency in the observer frame without any intrinsic dynamics. This behavior has no analogue in point-like systems and arises from the mapping of spatial structure into temporal variability under Lorentz transformation. 
{\color{black}
Unlike Doppler boosting, lighthouse effects, jet precession, shocks, turbulence, or magnetic reconnection, the proposed variability originates from Lorentz mapping of a stationary spatially structured magnetic field, with the characteristic frequency determined by the magnetic-field structure and the bulk Lorentz factor.
}
In this sense, the observed variability is directly determined by the intrinsic spatial structure of the field (through the helical wave number $k$), representing a nontrivial and previously unexplored consequence of relativistic transformations.

Observations of relativistic jets often reveal pronounced temporal variability while lacking sufficient spatial resolution to directly probe the underlying structure of the source. In such cases, such variability serves as a primary diagnostic of the physical conditions within the jet plasma flow and is typically attributed to intrinsic dynamical processes with corresponding temporal scales, including shocks, turbulence, magnetic reconnection, or geometric and Doppler effects. The results presented here demonstrate for the first time that this interpretation can be incomplete: even in the absence of intrinsic dynamics, a stationary, spatially structured magnetic field associated with a spatially extended source can produce observable time variability purely through relativistic effects. This behavior has no analogue in point-like systems and arises from the relativistic mapping of spatial structure into temporal variability.

{\color{black}

In addition to flux variability, Lorentz transformation of a stationary spatially structured magnetic field naturally produces temporal evolution of the observed polarization properties.  Lorentz mapping leads directly to variability of the Stokes parameters and polarization angle. Unlike the flux variability, which is intrinsically periodic, the polarization angle evolution is generally non-periodic. Consequently, a stationary magnetic structure may produce smooth polarization angle swings or rotations in the observer's frame purely as a result of relativistic projection effects, even though the underlying magnetic configuration remains stationary in the comoving frame. Interestingly, the polarization-angle variability depends strongly on the bulk Lorentz factor, introducing an observational bias toward rapid polarization angle swings in highly relativistic jets. 
}

Variability described in the present paper is naturally suited to high-frequency and 
quasi-periodic behavior in  relativistic jets, since the observed timescale is set by 
Lorentz mapping of spatial magnetic structure rather than by intrinsic plasma dynamics 
or light-crossing constraints. Consequently, very short observed timescales can 
arise in super-relativistic flows without requiring rapid physical evolution 
in the co-moving frame. 
While the present analysis employs a specific idealized helical (Beltrami) configuration, extensions to more general magnetic structures may enable inference of magnetic topology from observed variability.
These results identify a new relativistic effect in which 
Lorentz mapping of magnetic structure alone give rise to observable variability, 
providing a direct diagnostic of relativistic outflows.

{\color{black}

\appendix

\section{Derivation of the Observer Frame CK Field}

In this Appendix, we derive the observer-frame form of the Chandrasekhar–Kendall (CK) field starting from its stationary co-moving configuration. Following Eq.~(5), the Lorentz transformation matrix corresponding to the adopted symmetry is given by:
\begin{equation}
\Lambda^{\mu}{}_{\nu}(\boldsymbol{\beta}) =
\left[
\begin{array}{cccc}
\Gamma & 0 & -\Gamma \beta_y & -\Gamma \beta_z \\[4pt]
0 & 1 & 0 & 0 \\[4pt]
-\Gamma \beta_y & 0 &
\Gamma \frac{\beta_y^2}{\beta^2} + \frac{\beta_z^2}{\beta^2} &
(\Gamma - 1)\frac{\beta_y \beta_z}{\beta^2} \\[6pt]
-\Gamma \beta_z & 0 &
(\Gamma - 1)\frac{\beta_y \beta_z}{\beta^2} &
\Gamma \frac{\beta_z^2}{\beta^2} + \frac{\beta_y^2}{\beta^2}
\end{array}
\right] ~.
\end{equation}

In the considered case, the electric field is absent in the co-moving frame. Hence, the electromagnetic field tensor is determined entirely by the magnetic field components:
\begin{equation}
F^{0i}=0 ~,~~~
F^{ij}=-\epsilon_{ijk} B_k ~,
\end{equation}
where $\epsilon_{ijk}$ denotes the Levi--Civita antisymmetric symbol. Now electromagnetic field tensor calculation from Eq. (1) can be reduced to the following:
\begin{equation}
F^{ij}(x) = \Lambda^i{}_{\alpha}(-\boldsymbol{\beta}) \Lambda^j{}_{\beta}(-\boldsymbol{\beta})
F'^{\alpha\beta}(x') ~,
\end{equation}
Hence, the individual components of the magnetic field in the observer frame may be expressed in terms of their co-moving counterparts as follows:
\begin{eqnarray}
B_x(x) &=& \Gamma B'_x(x') ~, \\
B_y(x) &=& \left( \Gamma\cos^2\theta+\sin^2\theta \right)B'_y(x') -
(\Gamma-1)\sin\theta\cos\theta\,B'_z(x') ~, \\
B_z(x) &=& -(\Gamma-1)\sin\theta\cos\theta\,B'_y(x') + \left( \Gamma\sin^2\theta+\cos^2\theta
\right)B'_z(x') ~.
\end{eqnarray}

To obtain the magnetic-field structure in the observer frame, we employ the Chandrasekhar–Kendall (CK) mode projections from Eqs.~(15)–(19) and substitute them into Eqs.~(A.4)–(A.6), obtaining:
\begin{eqnarray}
B_x(x) &=& -\Gamma \frac{\alpha B_0}{2} \,\Big[  \cos(\theta) y^\prime + \sin(\theta) z^\prime \Big] \, \sin\! \left(\cos(\theta) k z^\prime  - \sin(\theta) k y^\prime\right) ~,
\\
B_y(x) &=& B_0 \Big[ \sin\theta + \Gamma\frac{\alpha}{2}\cos\theta\,x \Big] \sin\! \left( \cos(\theta) k z^\prime  - \sin(\theta) k y^\prime \right) ~,
\\
B_z(x) &=& B_0 \Big[ \cos\theta - \Gamma\frac{\alpha}{2}\sin\theta\,x \Big] 
\sin\! \left(\cos(\theta) k z^\prime  - \sin(\theta) k y^\prime\right) ~.
\end{eqnarray}

Since the magnetic field components are still defined in terms of the co-moving coordinates $x'^\mu$, the corresponding coordinate transformation must be applied to rewrite them in the observer frame coordinates $x^\mu$:
\begin{equation}
x'^{\mu} = \Lambda^\mu{}_{\eta}(\boldsymbol{\beta})x^\eta ~.
\end{equation}
Applying the Lorentz transformation matrix Eq.~(A.1) to the spacetime coordinates, we obtain:
\begin{eqnarray}
t' &=& \Gamma t - \Gamma\beta\sin\theta\, y - \Gamma\beta\cos\theta\, z ~,\\
x' &=& x ~, \\
y' &=& -\Gamma\beta\sin\theta\, ct + \left( \Gamma\sin^2\theta+\cos^2\theta \right) y 
+ (\Gamma-1)\sin\theta\cos\theta\, z ~, \\
z' &=&  -\Gamma\beta\cos\theta\,c t + (\Gamma-1)\sin\theta\cos\theta\, y + 
\left( \Gamma\cos^2\theta+\sin^2\theta \right)z ~.
\end{eqnarray}
Now, substituting Eqs.~(A.11)–(A.14) into Eqs.~(A.7)–(A.9), we finally obtain explicit expressions for the magnetic-field components as functions of the observer-frame spacetime coordinates, in the form:
\begin{eqnarray}
B_x({\bf r},t) &=& - B_0 {\Gamma \alpha \over 2} L_T({\bf r},t) \sin(\Phi_H({\bf r},t)) ~, \\
B_y({\bf r},t) &=& B_0 \left( \sin(\theta) + {\Gamma \alpha \over 2} \cos(\theta) x \right)
\sin(\Phi_H({\bf r},t)) ~, \\
B_z({\bf r},t) &=& B_0 \left( \cos(\theta) - {\Gamma \alpha \over 2} \sin(\theta) x \right)
\sin(\Phi_H({\bf r},t)) ~,
\end{eqnarray}
where we have introduced helical phase ($\Phi_H$) and transverse coordinate ($L_T$) defined in the observer's frame:
\begin{equation}
\Phi_H({\bf r},t) = (\Gamma_\theta - 1) \sin(\theta) k \, y + (\Gamma_\theta + 1) \cos(\theta) k \, z - \Gamma \cos(2\theta) k \beta \, ct  ~,
\end{equation}
\begin{equation}
L_T({\bf r},t) = (\Gamma - \Gamma_\theta) \cos(\theta) y + (\Gamma + \Gamma_\theta) \sin(\theta)  z - \Gamma \sin(2\theta) \beta \, ct  ~,
\end{equation}
and angle modified gamma factor as follows:
\begin{equation}
\Gamma_\theta \equiv (\Gamma-1) \cos(2 \theta) ~. 
\end{equation}

For completeness, we also calculate the electric field in the observer frame. Although the electric field vanishes in the co-moving frame, it is revealed in the observer frame through Lorentz mixing of the electric and magnetic field components:
\begin{equation}
\mathbf{E}(x)=-\Gamma\,\boldsymbol{\beta}\times \mathbf{B}'(x') .
\end{equation}
Using Eqs.~(A.1)–(A.3), and coordinate transformation (A.11)-(A.14) we derive:
\begin{eqnarray}
E_x(\mathbf r,t) &=& 
B_0 \frac{\Gamma \alpha}{2}\, \beta x\,\sin\!\left( \Phi_H(\mathbf r,t)\right) ~, \\
E_y(\mathbf r,t) &=& 
B_0 \frac{\Gamma \alpha}{2}\, \beta \cos\theta\,L_T(\mathbf r,t)\, \sin\!\left( \Phi_H(\mathbf r,t)\right) ~, \\
E_z(\mathbf r,t) &=& 
- B_0 \frac{\Gamma \alpha}{2}\, \beta \sin\theta\,L_T(\mathbf r,t)\, \sin\!\left( \Phi_H(\mathbf r,t)\right) ~.
\end{eqnarray}
Thus, Eqs.~(A.15)–(A.24) provide the complete observer frame electromagnetic structure corresponding to the stationary co-moving Chandrasekhar–Kendall (CK) configuration.

}

\section*{Acknowledgments}
{The work of VIB and NLS was partially supported by Shota Rustaveli
Georgian National Foundation Grant Project No. FR-22- 8273.}


\section*{References}
\begin{thebibliography}{99}

{\color{black}
\bibitem{Bromberg2024}
Bromberg O and Tchekhovskoy A 2024
``The formation, propagation and emission of relativistic jets''
Nature Astron. \textbf{8} 1060
}

\bibitem{BZ1977}
Blandford R D and Znajek R L 1977
``Electromagnetic extraction of energy from Kerr black holes''
Mon. Not. R. Astron. Soc. \textbf{179} 433

\bibitem{K2007}
Komissarov S S, Barkov M V, Vlahakis A V and McKinney J C 2007
``Magnetic acceleration of ultrarelativistic jets in gamma-ray burst sources''
Mon. Not. R. Astron. Soc. \textbf{380} 51

\bibitem{Blandford2019}
Blandford R D, Meier D G and Readhead A 2019
``Relativistic jets from active galactic nuclei''
Annu. Rev. Astron. Astrophys. \textbf{57} 467

\bibitem{R1966}
Rees M J 1966
``Appearance of relativistically expanding radio sources''
Nature \textbf{211} 468

\bibitem{C1977}
Cohen M H 1977
``Superluminal sources''
Annu. Rev. Astron. Astrophys. \textbf{15} 175

\bibitem{BK1979}
Blandford R D and K\"onigl A 1979
``Relativistic jets as compact radio sources''
Astrophys. J. \textbf{232} 34

\bibitem{BBR1984}
Begelman M C, Blandford R D and Rees M J 1984
``Theory of extragalactic radio sources''
Rev. Mod. Phys. \textbf{56} 255

\bibitem{LPG2005}
Lyutikov M, Pariev V I and Gabuzda D C 2005
``Polarization and structure of relativistic parsec-scale AGN jets''
Mon. Not. R. Astron. Soc. \textbf{360} 869

\bibitem{L2022}
Lyutikov M 2022
``On force-free magnetic knots''
Mon. Not. R. Astron. Soc. \textbf{517} 6177

\bibitem{Penrose1959}
Penrose R 1959
``The apparent shape of a relativistically moving sphere''
Proc. Camb. Philos. Soc. \textbf{55} 137

\bibitem{Terrell1959}
Terrell J 1959
``Invisibility of the Lorentz contraction''
Phys. Rev. \textbf{116} 1041

\bibitem{Hornof2025}
Hornof D et al 2025
``Experimental observation of the Penrose--Terrell effect''
Commun. Phys. \textbf{8} 161

\bibitem{CK1957}
Chandrasekhar S and Kendall P C 1957
``On force-free magnetic fields''
Astrophys. J. \textbf{126} 457

\bibitem{low}
Low B C 1982
``Nonlinear force-free magnetic fields''
Rev. Geophys. Space Phys. \textbf{20} 145

\bibitem{Woltier}
Woltjer L 1958
``A theorem on force-free magnetic fields''
Proc. Natl. Acad. Sci. U.S.A. \textbf{44} 489

\bibitem{Taylor}
Taylor J B 1986
``Relaxation and magnetic reconnection in plasmas''
Rev. Mod. Phys. \textbf{58} 741

\bibitem{kusano}
Kusano K, Suzuki Y and Nishikawa K 1995
``A solar flare trigger mechanism based on the Woltjer--Taylor minimum-energy state''
Astrophys. J. \textbf{441} 942

\bibitem{Moffatt}
Moffatt H K 1978
\textit{Magnetic Field Generation in Electrically Conducting Fluids}
(Cambridge: Cambridge University Press)

\bibitem{MY1998}
Mahajan S M and Yoshida Z 1998
``Double-curl Beltrami flow: Diamagnetic structures''
Phys. Rev. Lett. \textbf{81} 4863

\bibitem{mnsy}
Mahajan S M, Nikol’skaya K I, Shatashvili N L and Yoshida Z 2002
``Generation of astrophysical jets from magnetohydrodynamic equilibria''
Astrophys. J. \textbf{576} L161

\bibitem{osym}
Ohsaki S, Shatashvili N L, Yoshida Z and Mahajan S M 2001
``Thermodynamically supported emergence of astrophysical outflows''
Astrophys. J. \textbf{559} L61

\bibitem{RT1}
Yoshida Z et al 2010
``Self-organized helical states in extended magnetohydrodynamics''
Phys. Plasmas \textbf{17} 112507

\bibitem{SMB2T}
Shatashvili N L, Mahajan S M and Berezhiani V I 2019
``Formation of triple Beltrami states in degenerate electron--positron--ion plasmas''
Astrophys. Space Sci. \textbf{364} 148

\bibitem{SSMD}
Saralidze E, Shatashvili N L, Mahajan S M and Dadiani E 2025
``Generalized Beltrami equilibria and relativistic jets''
Astrophys. Space Sci. \textbf{370} 64

\bibitem{Katsadze}
Katsadze E, Revazashvili N and Shatashvili N L 2024
``Relativistic jets generated by generalized Beltrami equilibria''
J. High Energy Astrophys. \textbf{43} 20

{\color{black}

\bibitem{Mizuno2009}
Mizuno Y, Hardee P E and Nishikawa K I 2009
"Three-dimensional relativistic magnetohydrodynamic simulations of current-driven instability. I. Instability criterion and nonlinear evolution"
Astrophys. J. \textbf{696} 1492

\bibitem{Mizuno2012}
Mizuno Y, Hardee P E and Nishikawa K I 2012
"Three-dimensional relativistic magnetohydrodynamic simulations of current-driven instability. III. Rotating relativistic jets"
Astrophys. J. \textbf{757} 16

}



\bibitem{Y1991}
Yoshida Z 1991
``Self-organization in magnetohydrodynamic fluids''
Prog. Theor. Phys. \textbf{86} 474

\bibitem{GS1965}
Ginzburg V L and Syrovatskii S I 1965
``Cosmic magnetobremsstrahlung (synchrotron radiation)''
Annu. Rev. Astron. Astrophys. \textbf{3} 297

\bibitem{RL1979}
Rybicki G B and Lightman A P 1979
\textit{Radiative Processes in Astrophysics}
(New York: Wiley)


\bibitem{Zhang}
Zhang B 2024
``Fast radio bursts: progress, puzzles and prospects''
Phys. Rep. \textbf{1072} 1

\bibitem{Kim2024}
Kim D E, Di Gesu L, Liodakis I, Marscher A P, Jorstad S G, Middei R et al 2024
``Magnetic field properties inside the jet of Mrk 421: Multiwavelength polarimetry, including the Imaging X-ray Polarimetry Explorer''
Astron. Astrophys. \textbf{681} A12

\bibitem{Liodakis2025}
Liodakis I, Kiehlmann S, Marscher A P, Zhang H, Blinov D, Jorstad S G, Agudo I et al 2024
``Testing particle acceleration in blazar jets with continuous high-cadence optical polarization observations''
Astron. Astrophys. \textbf{689} A200

\bibitem{Albert2007}
Albert J et al 2007
``Variable very-high-energy gamma-ray emission from Markarian 501''
Astrophys. J. \textbf{669} 862

\bibitem{MAGIC}
MAGIC Collaboration 2020
``Unprecedented study of the broadband emission of Mrk 421 during flaring activity in March 2010''
Astron. Astrophys. \textbf{643} A14



\end{thebibliography}
\end{document}